\documentclass{ws-procs9x6}

\begin{document}

\title{TMDs and Drell-Yan Experiments at Fermilab and J-PARC}

\author{Jen-Chieh Peng}

\address{Department of Physics, University of Illinois at 
Urbana-Champaign\\
Urbana, IL 61801, U.S.A.\\
E-mail: jcpeng@illinois.edu}

\begin{abstract}
The roles of the Drell-Yan experiments in studying the 
Transverse-Momentum-Dependent (TMD) parton distributions are
discussed. Recent results from the Fermilab E866 experiment on
the angular distributions of Drell-Yan dimuons in $p+p$ and
$p+d$ at 800 GeV/c are presented. These data are
compared with the pion-induced Drell-Yan data,
and with models which attribute the $\cos 2 \phi$ azimuthal distribution
to the presence of the transverse-momentum-dependent Boer-Mulders
structure function $h_1^\perp$. Constraints on the magnitude of 
the sea-quark $h_1^\perp$ structure functions are obtained.
Future prospects for studying the TMDs with Drell-Yan experiments
at Fermilab and J-PARC are also discussed.
\end{abstract}


\bodymatter

\section{Introduction}\label{aba:sec1}
While our current knowledge on the partonic substructure of nucleons
is obtained largely from deep-inelastic scattering (DIS) experiments, the
Drell-Yan process~\cite{drell70}, in which a charged lepton pair 
is produced in a hadron-hadron interaction via the 
electromagnetic $q \bar q \to l^+ l^-$ process, also provides 
unique information on parton distributions. In particular, 
the Drell-Yan process has been used to determine the antiquark 
contents of nucleons and nuclei~\cite{pat99}, as well as
the quark distributions of pions, kaons, and antiprotons~\cite{kenyon82}. 
Such information is difficult, if not impossible, to obtain from 
DIS experiments. As the Drell-Yan process can be well 
described by next-to-leading order QCD calculations~\cite{stirling93}, 
a firm theoretical framework exists for utilizing the
Drell-Yan process to extract the parton distributions.

The study of the transverse momentum dependent (TMD) parton distributions
of the nucleon has received much attention in recent years
as it provides new perspectives on the hadron structure and 
QCD~\cite{barone02}. These novel TMDs can be extracted from semi-inclusive
deep-inelastic (SIDIS) scattering experiments. Recent measurements of 
the SIDIS by the HERMES~\cite{hermes05}
and COMPASS~\cite{compass05} collaborations
have shown clear evidence for the existence of the T-odd Sivers functions.
These data also allow the first determination~\cite{vogelsang05} of
the magnitude and flavor structure of the Sivers functions and the nucleon
transversity distributions.

The TMD and transversity parton distributions can also be probed in
Drell-Yan experiments. As pointed out~\cite{ralston79} 
long time ago, the double 
transverse spin asymmetry in polarized Drell-Yan, $A_{TT}$, is proportional
to the product of transversity distributions, $h_1(x_q)h_1(x_{\bar q})$.
The single transverse spin asymmetry, $A_N$, is sensitive to the 
Sivers function~\cite{sivers90}, $f^\perp_{1T}(x)$ of 
the polarized proton (beam or target).
Even unpolarized Drell-Yan experiments can be used to probe the TMD
distribution function, since the cos$2\phi$ azimuthal angular dependence
is proportional to the product of two Boer-Mulders functions~\cite{boer98}, 
$h^\perp_1(x_1) \bar h^\perp_1(x_2)$. A unique feature of the Drell-Yan 
process is that, unlike the SIDIS, no fragmentation functions are involved.
Therefore, the Drell-Yan process provides an entirely independent technique
for measuring the TMD functions. Furthermore, the proton-induced Drell-Yan
process is sensitive to the sea-quark TMDs and can lead to flavor separation
of TMDs when combined with the SIDIS data. Finally, the intriguing 
prediction~\cite{collins02} that the T-odd TMDs extracted from 
DIS will have a sign-change for the
Drell-Yan process remains to be tested experimentally.

In this article, we first present the measurement of Drell-Yan
azimuthal angular distributions in the Fermilab experiment E866. 
Implication of the data on the TMD Boer-Mulders function is discussed. 
Prospects for future Drell-Yan experiments for studying TMD and 
transversity distributions at Fermilab and J-PARC will also be discussed.

\section{Fermilab dimuon experiments and TMDs}\label{aba:sec2}

\subsection{Overview}
During the last two decades, a series of fixed-target dimuon
production experiments (E772, E789, E866) have been carried out
using 800 GeV/c proton beam at Fermilab.
At 800 GeV/c, the dimuon data contain Drell-Yan continuum up to
dimuon mass of $\sim 15$ GeV as well as quarkonium (J/$\Psi$, 
$\Psi^\prime$, and $\Upsilon$ resonances) productions. The Drell-Yan
process and quarkonium productions often provide complementary
information, since Drell-Yan is an electromagnetic process
via quark-antiquark annihilation while the quarkonium production
is a strong interaction process dominated by gluon-gluon fusion
at this beam energy.
 
The Fermilab dimuon experiments cover a broad range of physics
topics. The Drell-Yan data have provided informations on the
antiquark distributions in the nucleons~\cite{pat92,hawker98,
peng98,towell01} and nuclei~\cite{alde90,vasiliev99}.
These results showed the surprising results that the
antiquark distributions in the
nuclei are not enhanced~\cite{alde90,vasiliev99},
contrary to the predictions of models which explain the
EMC effect in term of nuclear
enhancement of exchanged mesons. Moreover, the Drell-Yan cross
section ratios $(p+d)/(p+p)$ clearly establish the
flavor asymmetry of the $\bar d$ and $\bar u$ distributions in
the proton, and they map out the $x$-dependence of this
asymmetry~\cite{hawker98,peng98,towell01}.
A recent analysis of the $\Upsilon$ production cross section ratios for
$(p+d)/(p+p)$ has shown that the gluon distributions in proton and neutron
are very similar~\cite{zhu08}.
Pronounced nuclear dependences of quarkonium productions as a function
of $x_F$ and $p_T$ have
been observed for J/$\Psi$, $\Psi^\prime$, and $\Upsilon$ resonances
~\cite{alde91a,alde91b,kowitt94,leitch95}. The nuclear Drell-Yan cross
sections also exhibit $x_F$ as well as $p_T$ dependences~\cite{vasiliev99,
pat99,johnson01,johnson02}, which are weaker than the quarkonium
nuclear dependences but can provide information on
the energy loss of quarks traversing the nucleus~\cite{garvey03}.

Several review articles covering some of these results are
available~\cite{pat99,garvey01,reimer07}. In the following, we
will focus on the recent results from experiment E866 on the
measurement of angular distributions of the Drell-Yan cross
sections.

\subsection{Angular distributions of Drell-Yan and the Boer-Mulders function}

Despite the success of perturbative QCD in describing the Drell-Yan cross
sections, it remains a challenge to understand the angular
distributions of the Drell-Yan process. Assuming dominance of the
single-photon process, a general expression for the Drell-Yan
angular distribution is~\cite{lam78}
\begin{equation}
\frac {d\sigma} {d\Omega} \propto 1+\lambda \cos^2\theta +\mu \sin2\theta
\cos \phi + \frac {\nu}{2} \sin^2\theta \cos 2\phi,
\label{eq:eq1}
\end{equation}
\noindent where $\theta$ and $\phi$ denote the polar and azimuthal angle,
respectively, of the $l^+$ in the dilepton rest frame. In the ``naive"
Drell-Yan model, where the transverse momentum of the quark is ignored
and no gluon emission is considered, $\lambda =1$ and $\mu = \nu =0$ are
obtained. QCD effects~\cite{chiappetta86} and
non-zero intrinsic transverse momentum of the quarks~\cite{cleymans81}
can both lead to $\lambda \ne 1$ and $\mu, \nu \ne 0$. However,
$\lambda$ and $\nu$ should still
satisfy the relation
$1-\lambda = 2 \nu$~\cite{lam78}. This so-called Lam-Tung relation, obtained as
a consequence of the spin-1/2 nature of the quarks, is analogous
to the Callan-Gross relation~\cite{callan69}
in deep-inelastic scattering.

The first measurement of the Drell-Yan angular distribution was
performed by the NA10 Collaboration for $\pi^- + W$ with the
highest statistics at 194
GeV/c~\cite{falciano86,guanziroli88}.
The $\cos 2 \phi$ angular dependences showed a sizable $\nu$,
increasing with dimuon transverse momentum ($p_T$) and reaching a value
of $\approx 0.3$ at $p_T = 2.5$ GeV/c. The Fermilab E615 Collaboration
subsequently performed a measurement of $\pi^- + W$ Drell-Yan production
at 252 GeV/c with broad coverage in the
decay angle $\theta$~\cite{conway89}. The E615 data showed that
the Lam-Tung relation, $2\nu = 1 - \lambda$, is clearly violated.

The NA10 and E615 results on the Drell-Yan angular distributions strongly
suggest that new effects beyond conventional perturbative QCD are present.
Brandenburg, Nachtmann and Mirke suggested that a factorization-breaking
QCD vacuum may lead to a
correlation between the transverse spin of the antiquark in the pion and
that of the quark in the nucleon~\cite{brandenburg93}.
This would result in a non-zero
$\cos 2\phi$ angular dependence consistent with the data.
Several authors
have also considered higher-twist effects from quark-antiquark binding
in pions~\cite{brandenburg94,eskola94}.
However, the model is strictly applicable
only in the $x_\pi \to 1$ region, while the NA10 and E615 data exhibit
nonperturbative effects over a much broader kinematic region.

More recently, Boer pointed out~\cite{boer99}
that the $\cos 2 \phi$ angular dependences
observed in NA10 and E615 could be due to the $k_T$-dependent
parton distribution function $h_1^\perp$. This so-called Boer-Mulders
function~\cite{boer98} is an example of a novel type of $k_T$-dependent
parton distribution function, and it
characterizes the correlation of a quark's transverse spin and
its transverse momentum, $k_T$, in an unpolarized nucleon.
The Boer-Mulders function is the
chiral-odd analog of the Sivers function and
owes its existence to the presence of initial/final state
interactions~\cite{boer03}. While the Sivers function is
beginning to be quantitatively determined from
the SIDIS experiments, very little is known about the Boer-Mulders
function so far.

Several model calculations have been carried out for the
Boer-Mulders functions. In the quark-diquark model, it was
shown that the Boer-Mulders functions are identical to
the Sivers functions when only the scalar diquark configuration
is considered~\cite{boer03, gamberg03}. More recently, calculations
taking into account both
the scalar and the axial-vector diquark configurations found significant
differences between the Sivers and Boer-Mulders
functions~\cite{gamberg08}.
In particular, the $u$ and $d$ quark Boer-Mulders functions are predicted
to be both negative, while the Sivers function is negative for the $u$
quark and positive for the $d$ quark. Other calculations
using the MIT bag model~\cite{yuan03}, the relativistic constituent
quark model~\cite{pasquini07}, the large-$N_c$ model~\cite{pobylitsa03},
and the lattice QCD~\cite{gockeler07}
also predicted negative signs for the $u$ and $d$ Boer-Mulders
functions. Burkardt recently pointed out~\cite{burkardt08}
that the origin of the
negative signs for the Boer-Mulders functions lies in the
phase of the $p$-wave lower component relative to the
$s$-wave upper component of the quark
wave function for the solution to the free Dirac
equation. The model predictions
for the same signs of the $u$ and $d$  Boer-Mulders functions
remain to be tested experimentally.

\subsection{Results from Fermilab E866}

To shed additional light on the origins of the NA10 and E615 Drell-Yan
angular distributions, we recently analyzed $p+p$ and $p+d$ Drell-Yan angular
distribution data at 800 GeV/c from Fermilab E866.
There had been no report on
the azimuthal angular distributions for proton-induced Drell-Yan -- all
measurements were for polar angular
distributions~\cite{pat99,chang03,brown01}.
Proton-induced Drell-Yan data provide a test of
theoretical models. For example, the $\cos 2\phi$ dependence is expected
to be much reduced in proton-induced Drell-Yan if the underlying mechanism
involves the Boer-Mulders functions. This is due to the expectation that
the Boer-Mulders functions are small for the sea-quarks~\cite{lu07}. 
However, if the
QCD vacuum effect~\cite{brandenburg93} is the origin of
the $\cos 2 \phi$ angular dependence, then the azimuthal behavior of
proton-induced Drell-Yan should be similar to that of pion-induced
Drell-Yan.
Finally, the validity of the Lam-Tung relation has never been
tested for proton-induced Drell-Yan, and the present
study provides a first test.

The Fermilab E866 experiment was performed using the upgraded Meson-East
magnetic pair spectrometer. An 800 GeV/c primary proton
beam with up to $2 \times 10^{12}$ protons per 20-second beam spill
was incident upon 50.8 cm long target flask containing either liquid
hydrogen, liquid deuterium or vacuum.
The detector system consisted of
four tracking stations and a momentum analyzing magnet.
From the momenta of the $\mu^+$ and $\mu^-$, kinematic variables of
the dimuons ($x_F, m_{\mu\mu}, p_T$) were readily reconstructed.
The muon angles $\theta$ and $\phi$ in the Collins-Soper
frame~\cite{collins77} were also calculated. To remove the
quarkonium background,
only events with $4.5 <m_{\mu\mu}<
9$ GeV/c$^2$ or $m_{\mu\mu} > 10.7$ GeV/c$^2$ were analyzed.
A total of $\sim$54,000 $p+p$ and $\sim$118,000 p+d Drell-Yan 
events covering the decay angular range $-0.5 < \cos\theta
<0.5$ and $-\pi < \phi < \pi$ remain.

The measurement of the $\cos 2 \phi$ dependence for the $p+d$ 
Drell-Yan cross sections has already been reported~\cite{zhu07}. 
As shown in Fig. 1, significantly smaller (but non-zero) cos$2\phi$ 
azimuthal angular dependence was observed in the $p+d$ reaction.
While pion-induced Drell-Yan process is dominated by annihilation 
between a valence antiquark in the pion and a valence quark in the 
nucleon, proton-induced Drell-Yan process involves a valence quark 
annihilating with a sea antiquark in the nucleon.
Therefore, the $p+d$ result suggests~\cite{zhu07} that the Boer-Mulders 
functions for sea antiquarks are significantly smaller than those for 
valence quarks.

\begin{figure}
\begin{center}
\psfig{file=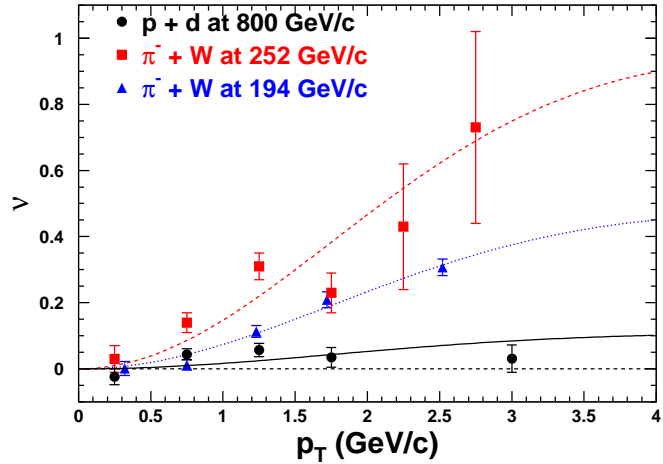,width=3.5in}
\end{center}
\caption{Parameter $\nu$ vs.\ $p_T$ in the Collins-Soper
frame for three Drell-Yan measurements. Fits to the data 
using an empirical expression~\cite{boer99} are also shown.}
\label{aba:fig1}
\end{figure}

A recent analysis~\cite{zhang08} of the $p+d$ $\cos 2 \phi$ data
showed that the sea-quark Boer-Mulders functions are indeed smaller
by a factor $\sim 5$ than the valence-quark Boer-Mulders functions.
This analysis also indicated that the E866 $p+d$
data are consistent with the $u$ and $d$ Boer-Mulders
functions having the same signs, as predicted by various models. However, the
$p+d$ data alone can not provide an unambiguous determination of the
flavor dependence of the Boer-Mulders functions. Moreover, it was
recently pointed out~\cite{boer06,berger07} that QCD 
processes would lead to sizeable $\cos 2\phi$
effect which has not been taken into account in the
extractions~\cite{boer99,zhang08,zhang08a} of Boer-Mulders
functions from the Drell-Yan data. The $p+p$ Drell-Yan data 
should provide further constraints on
the flavor dependence of the Boer-Mulders functions~\cite{zhang08a}.
It is also interesting to compare the 
$p+p$ and $p+d$ data with the prediction of QCD.

Figure 2 shows the angular distribution parameters $\lambda, \mu,$ and
$\nu$ vs.\ $p_T$ for the E866 $p+p$ and $p+d$ Drell-Yan data.
Within statistics, the angular distributions of $p+p$ are consistent
with those of $p+d$. Also shown in Fig. 2 is the quantity
$2\nu -(1-\lambda)$, which should vanish if the Lam-Tung relation is
valid. Figure 2 shows that the Lam-Tung relation
is indeed quite well satisfied within statistics. This is different
from the observation of
a significant violation of the Lam-Tung relation by the E615 collabortaion
in the $\pi^- + W$ reaction at 252 GeV/c~\cite{conway89}.

\begin{figure}
\begin{center}
\psfig{file=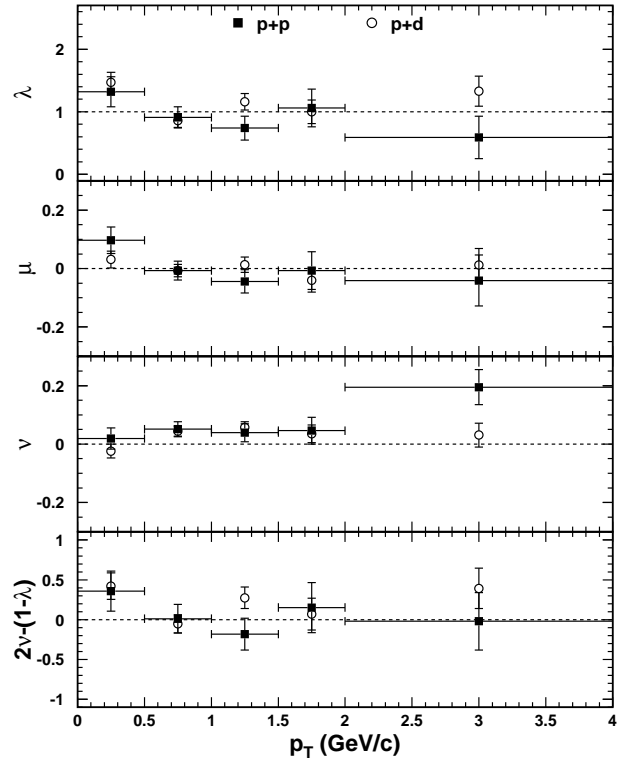,width=3.5in}
\end{center}
\caption{ Parameters $\lambda, \mu, \nu$ and $2\nu - (1-\lambda)$
vs.\ $p_T$ in the Collins-Soper frame. Solid squares (open circles)
are for E866 $p+p$ ($p+d$) at 800 GeV/c.}
\end{figure}

Figure 3 shows the parameter $\nu$ vs.\ $p_T$ for the $p+p$ and 
$p+d$ Drell-Yan data. The solid curves are 
calculations~\cite{zhang08,zhang08a} for $p+p$ and $p+d$ using 
parametrizations based on a fit to the $p+d$ Drell-Yan data. 
The larger values of $\nu$ for $p+p$ compared to $p+d$ are in 
qualitative agreement with the prediction. However, the shape of 
the predicted $p_T$ dependence is different
from the data. This strongly suggests that there could
be other mechanisms contributing to the $\cos 2\phi$ azimuthal angular
dependence at large $p_T$ region. In a recent paper~\cite{berger07}, 
the QCD contribution
to the $\cos 2\phi$ azimuthal angular dependence is given as
\begin{equation}
\nu = \frac {Q^2_\perp/Q^2} {1+\frac{3}{2}Q^2_\perp/Q^2},
\label{eq:eq2}
\end{equation}
\noindent where $Q_\perp$ is the dimuon transverse momentum. The 
predicted QCD contribution is shown as the dotted curve is Fig. 3.
Although neither the prediction based on Boer-Mulders functions
nor the QCD prediction could describe the entire range of $p_T$,
it is evident that the QCD contribution becomes important 
at high $p_T$ and the
Boer-Mulders functions contribute at lower $p_T$. 
An analysis combining both effects is required before a reliable 
extraction of the Boer-Mulders functions could be obtained.

\begin{figure}
\begin{center}
\psfig{file=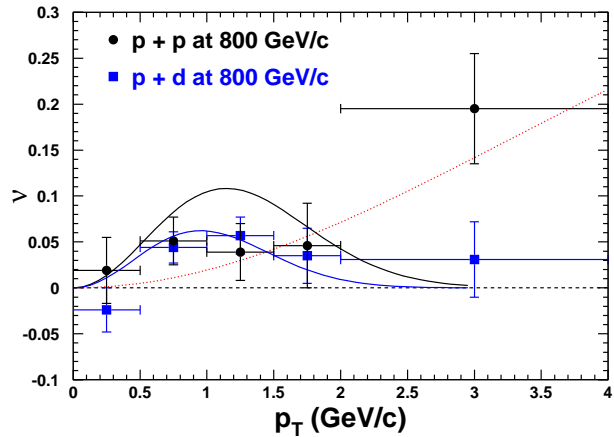,width=3.5in}
\end{center}
\caption{Parameter $\nu$ vs.\ $p_T$ for the $p+p$ and $p+d$ Drell-Yan data. 
The solid curves are
calculations~\cite{zhang08} for $p+p$ and $p+d$ using parametrizations
based on a fit to the $p+d$ Drell-Yan data. The dotted curve is the
contribition from QCD process (Eq. 2).}
\label{aba:fig3}
\end{figure}

\section{Future Prospects at Fermilab and J-PARC}

Future fixed-target dimuon experiments have been proposed at
the 120 GeV Fermilab Main Injector (FMI) and the 50 GeV J-PARC
facilities. The Fermilab proposal~\cite{e906}, E906, has been
approved and is expected to start data-taking around 2011. Two
dimuon proposals (P04~\cite{p04} and P24~\cite{p24}) have also
been submitted to the J-PARC for approval. The lower beam energies
at FMI and J-PARC present opportunities for extending the Drell-Yan
measurements to larger
$x$ ($x>0.25$). For given values of $x_1$ and $x_2$, the Drell-Yan cross
section is proportional to $1/s$, hence a gain of $\sim 16$ times in the
Drell-Yan cross sections can be obtained at the J-PARC energy of 50 GeV.
This would extend the measurement of Boer-Mulders functions to
larger $x$. In addition, the nuclear dependence of the Boer-Mulders 
functions can also be measured. 

The dimuon physics program at J-PARC is proposed to be carried out in
several stages. Since 30 GeV proton beam will be available at the initial
phase of J-PARC, the first measurements will focus on $J/\Psi$ production
at 30 GeV. This will be followed by measurements of Drell-Yan and
quarkonium production at 50 GeV after the beam energy is upgraded to
50 GeV. Experiments using polarized target could already be performed
with unpolarized beams. When polarized proton beam becomes available at
J-PARC, a rich and unique program on spin physics could also be pursued
at J-PARC using the dimuon spectrometer.

An important feature of $J/\Psi$ production using 30 or 50 GeV proton
beam is the dominance of the quark-antiquark annihilation subprocess.
This is in striking contrast to $J/\Psi$ production at 800 GeV (Fermilan
E866) or at 120 GeV (Fermilab E906), where the gluon-gluon fusion is the
dominant process. This suggests an exciting opportunity to use $J/\Psi$
production at J-PARC as an alternative method to probe antiquark
distribution.

With the possibility to accelerate polarized proton beams at J-PARC,
the spin structure of the proton can also be investigated with the
proposed dimuon experiments. In particular, polarzied Drell-Yan process
with polarized beam and/or polarized target at J-PARC would allow a unique
program on spin physics complementary to polarized DIS experiments
and the RHIC-Spin program. Specific physics topics include the measurements
of T-odd Boer-Mulders distribution function in unpolarized Drell-Yan,
the extraction of T-odd Sivers distribution functions in singly
transversely polarized Drell-Yan, the helicity distribution of
antiqaurks in doubly longitudinally polarized Drell-Yan, and the
transversity distribution in doubly transversely polarized
Drell-Yan. It is worth
noting that polarized Drell-Yan is one of the major physics program at the
GSI Polarized Antiproton Experiment (PAX). The RHIC-Spin program will
likely provide the first results on polarized Drell-Yan. However, the
high luminosity and the broad kinematic coverage for the
large-$x$ region at J-PARC would allow some unique measurements to be
performed in the J-PARC dimuon experiments.

\section*{Acknowledgments}

I am very grateful to Lingyan Zhu, Paul Reimer, and my collaborators 
on the E772, E789, and E866 experiments at Fermilab. Illuminating comments
from Matthias Burkardt, Feng Yuan, Bo-Qiang Ma, Werner Vogelsang, and 
Jianwei Qiu are greatly appreciated. I would also like to 
acknowledge the collaboration with Shin'ya Sawada and Yuji Goto 
on the J-PARC dimuon proposals.

\end{document}